\documentclass[twocolumn]{aastex}
\usepackage[english]{babel}
\usepackage{amsmath}
\usepackage{graphicx}
\usepackage{wasysym}
\usepackage{soul}
\usepackage{parskip}
\usepackage{wrapfig}
\usepackage{url}
\usepackage{float}
\usepackage{hyperref}
\usepackage{xcolor}

\newcommand{\tess}{\emph{TESS}}

\colorlet{Mycolor1}{green!72!red!228!}

\shorttitle{Gravity-Darkening Analysis of MASCARA-4 b}
\shortauthors{Ahlers et al.}
\graphicspath{{./}{figures/}}

\begin{document}

\title{Gravity-Darkening Analysis of Misaligned Hot Jupiter MASCARA-4 b}

\correspondingauthor{John P. Ahlers}
\email{johnathon.ahlers@nasa.gov}

\author{John P. Ahlers}
\affiliation{Exoplanets and Stellar Astrophysics Laboratory, Code 667, NASA Goddard Space Flight Center, Greenbelt, MD 20771, USA}

\author{Ethan Kruse}
\affiliation{Exoplanets and Stellar Astrophysics Laboratory, Code 667, NASA Goddard Space Flight Center, Greenbelt, MD 20771, USA}

\author{Knicole D. Col\'on}
\affiliation{Exoplanets and Stellar Astrophysics Laboratory, Code 667, NASA Goddard Space Flight Center, Greenbelt, MD 20771, USA}

\author{Patrick Dorval}
\affiliation{Leiden Observatory, Leiden University, Postbus 9513, 2300 RA Leiden, The Netherlands}
\affiliation{NOVA Optical IR Instrumentation Group at ASTRON, P.O. Box 2, 7990 AA Dwingeloo, The Netherlands}

\author{Geert Jan Talens}
\affiliation{Institut de Recherche sur les Exoplan\`etes, D\'{e}partement de Physique, Universit\'{e} de Montr\'{e}al, Montr\'{e}al, QC H3C 3J7, Canada}

\author{Ignas Snellen}
\affiliation{Leiden Observatory, Leiden University, Postbus 9513, 2300 RA Leiden, The Netherlands}

\author{Simon Albrecht}
\affiliation{Stellar Astrophysics Centre (SAC), Department of Physics and Astronomy, Aarhus University, Ny Munkegade 120, DK-8000 Aarhus C, Denmark}

\author{Gilles Otten}
\affiliation{Aix Marseille Univ, CNRS, CNES, LAM, Marseille, France}

\author{George Ricker}
\affiliation{Department of Physics and Kavli Institute for Astrophysics and Space Research, Massachusetts Institute of Technology, Cambridge, MA 02139, USA}

\author{Roland Vanderspek}
\affiliation{Department of Physics and Kavli Institute for Astrophysics and Space Research, Massachusetts Institute of Technology, Cambridge, MA 02139, USA}

\author{David Latham}
\affiliation{Harvard-Smithsonian Center for Astrophysics, 60 Garden Street, Cambridge, MA 02138, USA}

\author{Sara Seager}
\affiliation{Department of Physics and Kavli Institute for Astrophysics and Space Research, Massachusetts Institute of Technology, Cambridge, MA 02139, USA}
\affiliation{Department of Earth, Atmospheric and Planetary Sciences, MIT, Cambridge, MA 02139, USA}
\affiliation{Department of Aeronautics and Astronautics, MIT, Cambridge, MA 02139, USA}

\author{Joshua Winn}
\affiliation{Department of Astrophysical Sciences, Princeton University, Princeton, NJ 08544, USA}

\author{Jon M. Jenkins}
\affiliation{NASA Ames Research Center, Moffett Field, CA 94035, USA}


\author{Kari Haworth}
\affiliation{Department of Physics and Kavli Institute for Astrophysics and Space Research, Massachusetts Institute of Technology, Cambridge, MA 02139, USA}

\author{Scott Cartwright}
\affiliation{Proto-Logic LLC, 1718 Euclid Street NW, Washington, DC 20009, USA}

\author{Robert Morris}
\affiliation{NASA Ames Research Center, Moffett Field, CA 94035, USA}
\affiliation{SETI Institute, 189 Bernardo Avenue, Suite 200, Mountain View, CA 94043, USA}

\author{Pam Rowden}
\affiliation{School of Physical Sciences, The Open University, Milton Keynes MK7 6AA, UK}

\author{Peter Tenenbaum}
\affiliation{NASA Ames Research Center, Moffett Field, CA 94035, USA}
\affiliation{SETI Institute, 189 Bernardo Avenue, Suite 200, Mountain View, CA 94043, USA}


\author{Eric B. Ting}
\affiliation{NASA Ames Research Center, Moffett Field, CA 94035, USA}

\begin{abstract}
MASCARA-4 b is a hot Jupiter in a highly-misaligned orbit around a rapidly-rotating A3V star that was observed for 54 days by the Transiting Exoplanet Survey Satellite (\tess). We perform two analyses of MASCARA-4 b using a stellar gravity-darkened model. First, we measure MASCARA-4 b's misaligned orbital configuration by modeling its \tess~photometric light curve. We take advantage of the asymmetry in MASCARA-4 b's transit due to its host star's gravity-darkened surface to measure MASCARA-4 b's true spin-orbit angle to be $104^{\circ+7^\circ}_{-13^\circ}$. We also detect a $\sim4\sigma$ secondary eclipse at $0.491\pm0.007$ orbital phase, proving that the orbit is slightly eccentric. Second, we model MASCARA-4 b's insolation including gravity-darkening and find that the planet's received XUV flux varies by $4$\% throughout its orbit. MASCARA-4 b's short-period, polar orbit suggests that the planet likely underwent dramatic orbital evolution to end up in its present-day configuration and that it receives a varying stellar irradiance that perpetually forces the planet out of thermal equilibrium. These findings make MASCARA-4 b an excellent target for follow-up characterization to better understand orbital evolution and current-day  of planets around high-mass stars.
\end{abstract}


\section{Introduction} \label{sec:intro}
MASCARA-4 b (bRing-1 b) is a hot Jupiter in a highly spin-orbit misaligned 2.82 day orbit around the bright ($V_\mathrm{mag}=8.19$) AV3 star HD 85628 (TIC 371443216). The transiting planet was first discovered with the MASCARA and bRing ground-based telescopes \citep{dorval2019mascara}, and was observed in sectors 10 and 11 of \tess's full frame images (FFIs) at 30-minute cadence \citep{2015JATIS...1a4003R} . \citet{dorval2019mascara} spectroscopically determined the mass of the planet to be $3.1 \pm 0.9M_\mathrm{Jup}$. The host star of this system, HD 85628, rotates with $v\sin(i)=46.5\pm1$ km/s \citep{dorval2019mascara}. Its high rotation flattens the star into an oblate shape and produces a pole-to-equator luminosity gradient -- called gravity-darkening -- brought about by its lowered equatorial effective temperature \citep{von1924radiative}. In this work we constrain MASCARA-4 b's orbit geometry and insolation including the gravity-darkening effect.

MASCARA-4 b is dynamically interesting because it orbits its host star in a nearly polar orbit. Some dynamic mechanism must have tilted either the planet's orbit, the host star's rotation axis, or the plane of the protoplanetary disk. \citet{dorval2019mascara} previously measured the planet's projected obliquity to be $247.5^{\circ+1.5^\circ}_{-1.7^\circ}$ via Doppler tomography. We apply the gravity-darkening technique \citep{barnes2009transit,2011ApJS..197...10B,ahlers2014,ahlers2015spin,masuda2015spin,barnes2015probable,ahlers2019dealing,zhou2019two} to \tess~photometry to further constrain the planet's orbit geometry and measure its true spin-orbit angle. Our results match previous observations that close-in giant planets around high-mass stars commonly have misaligned orbits \citep[e.g.,][]{winn2010hot,schlaufman2010evidence}. Additionally, MASCARA-4 b likely migrated inward to its 2.82-day orbit \citep[e.g.,][]{dawson2014tidal,petrovich2015hot}. A plausible scenario for the planet's orbital evolution is therefore dynamic scattering or resonance that increased both orbital eccentricity and inclination, and then tidal recircularization pulled the planet into its ultra-short-period, highly misaligned present-day configuration \citep{fabrycky2007shrinking,socrates2012super}. However, the evolution pathway of MASCARA-4 b merits further investigation. 

Ultimately, MASCARA-4 b resides in an environment that cannot occur around lower-mass stars. With a spin-orbit misaligned orbit around a rapidly-rotating, gravity-darkened star, MASCARA-4 b's exposure to the star's hotter poles and cooler equator varies throughout its orbit. Only stars above the Kraft break ($M_\star\geq1.3M_\odot$) are expected to maintain a high rotation rate throughout their lifetimes \citep{kraft1967studies,maeder2008physics}, so varying irradiance due to gravity-darkening likely does not occur around Sun-like and smaller stars after zero age main sequence. Additionally, spin-orbit misalignment appears to occur commonly around A/F stars; therefore this scenario of varying irradiance, which we call gravity-darkened seasons, may occur to a significant fraction of planets orbiting high-mass stars.

MASCARA-4 b is a useful test case for understanding planet formation and evolution. The results of this research directly address two outstanding questions in exoplanetary science: why do hot Jupiters exist, and why does spin-orbit misalignment occur? As a misaligned hot Jupiter, MASCARA-4 b likely underwent significant orbital evolution to get to its current configuration. We explain our methods for analyzing this interesting system in \S\ref{sec:methods}, we show our results from photometry and insolation modeling in \S\ref{sec:results}, and we discuss possible formation, evolution, and current-day processes of MASCARA-4 b in \S\ref{sec:discussion}.

\section{Methods}\label{sec:methods}
We model the gravity-darkening effect on MASCARA-4 b in two ways. First, we model MASCARA-4 b's \tess~photometric light curve with gravity-darkening to determine the planet's orbit geometry. Second, we model the planet's insolation to show how its received stellar flux is influenced by stellar gravity-darkening. We discuss both approaches in the following subsections.

\subsection{\emph{TESS} Photometry}\label{sec:methods.tess}
\subsubsection{Data Processing}

\begin{figure*}
\includegraphics[width=\textwidth]{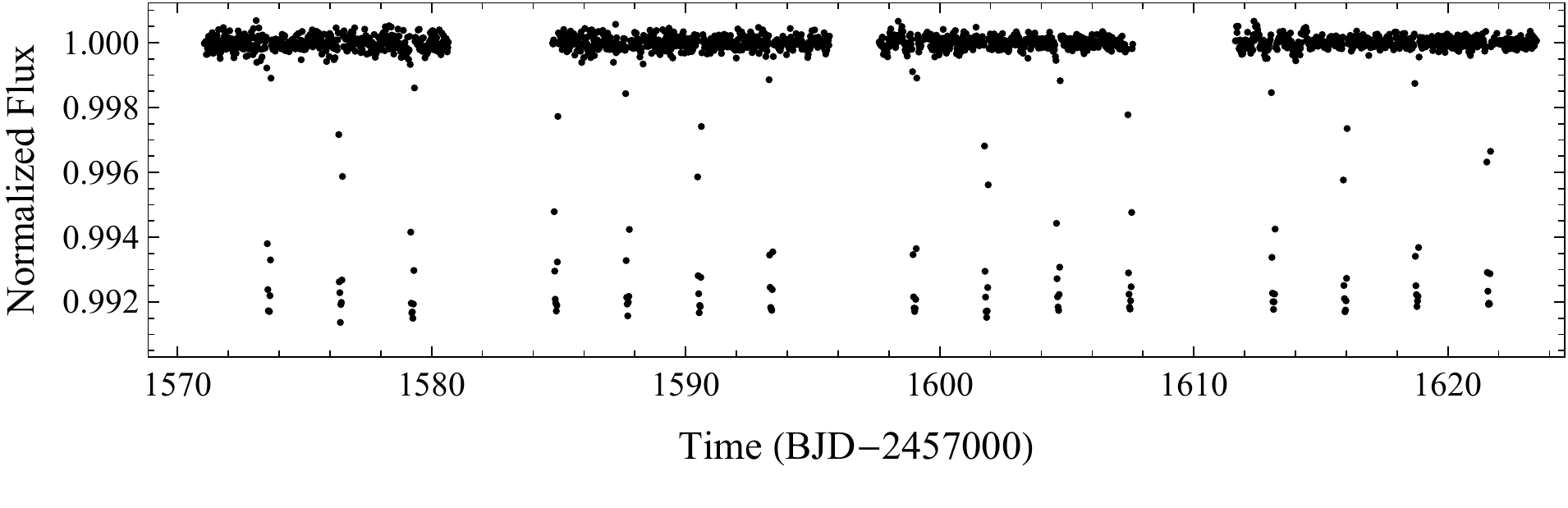}
\caption{\footnotesize \tess~observed 16 transit events of MASCARA-4 b during its sectors 10 and 11 observing campaigns at 30-minute cadence. We remove the first transit of Sector 10 from our dataset -- which was contaminated by excess scattered light -- leaving 15 transits for our analysis. We show the detrended MASCARA-4 b \tess~light curve here.}
\label{fig:tsplot}
\end{figure*}

Sixteen transits of MASCARA-4 b (TIC 371443216) were observed in \tess's Full Frame Images (FFI) at 30-minute cadence in sectors 10 and 11 from March 26, 2019 to May 21, 2019 during the southern observing campaign.  The FFIs were produced by the Science Processing Operations Center (SPOC) at NASA Ames Research Center \citep{jenkinsSPOC2016} and downlinked from the Mikulski Archive for Space Telescopes.  We create light curves using \texttt{eleanor} version 0.2.7 \citep{feinstein2019}. From \texttt{eleanor}'s various reduction options, we choose to use the point-spread-function-modeled light curve because it has the least noise on transit timescales.

The available \tess~photometry of MASCARA-4 b is broken up into four 13.5-day segments due to \tess's orbit. The first day of sector 10 was contaminated by large amounts of scattered light, increasing the noise and making transit analysis difficult. We remove this first day, which contained the first transit, leaving 15 transits used in this work. We apply a 15-hour moving average to the out-of-transit flux to correct for long-term systematics in each segment, normalizing the light curve to 1.0. We show the full normalized light curve in Figure \ref{fig:tsplot}. We phase-fold the light curve on MASCARA-4 b's orbital period and re-bin at 120 seconds to reduce computation time, following previous gravity-darkening works \citep{2011ApJS..197...10B,ahlers2014,ahlers2015spin,masuda2015spin,barnes2015probable,ahlers2019dealing}.

\begin{figure}[htbp]
\centering
\includegraphics[width=0.46\textwidth]{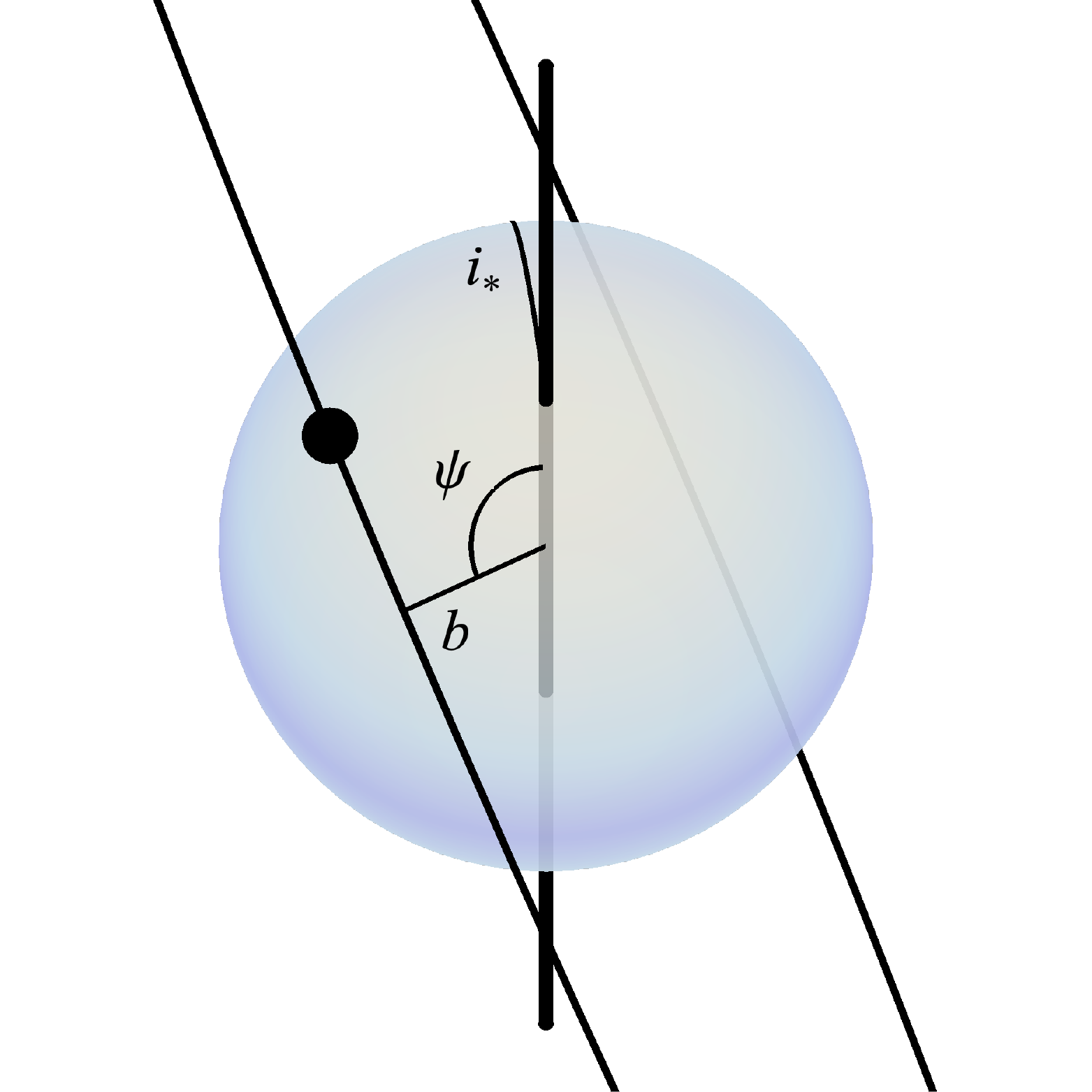}
\caption{\footnotesize The gravity-darkening technique measures three point-of-view orbit geometry parameters that together yield the true spin-orbit angle. The stellar inclination ($i_\star$) is the star's rotation axis tilt toward/away the viewer. The projected obliquity ($\psi$) is the projected tilt of the planet's orbit in the plane of the sky, and is the same angle measured by Doppler tomography. The orbital inclination ($i$) is the planet's orbital tilt toward/away the viewer, and is defined by $\cos(i)=bR_\star/r$, where $b$ is the impact parameter and $r$ is the planet's distance from the star. The star's color gradient represents its gravity-darkened surface.}
\label{fig:angles}
\end{figure}

\subsubsection{Transit Fitting and Gravity-Darkening}\label{methods.fit}

\begin{figure}[htbp]
\includegraphics[width=0.47\textwidth]{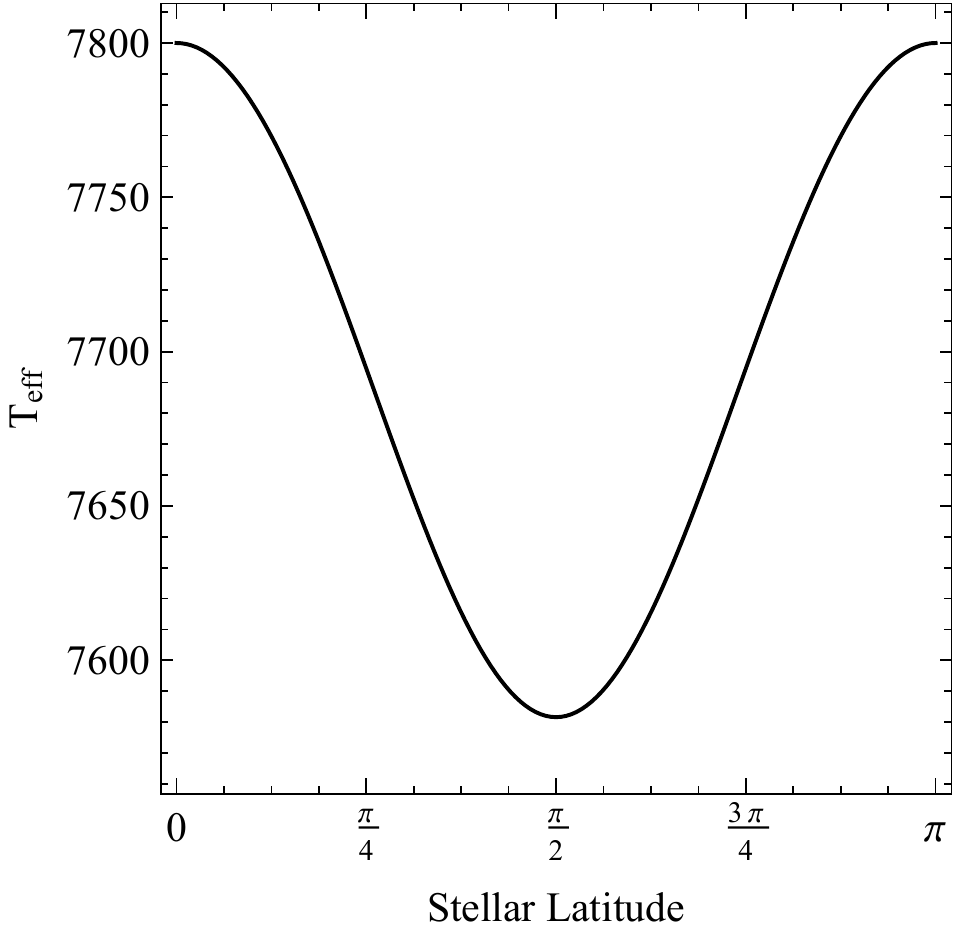}
\caption{\footnotesize The MASCARA-4 b host star varies roughly 220 K between its poles and equator due to its rapid rotation. Its $\sim3$\% change in local effective temperature corresponds to a $\sim12$\% change in local brightness, resulting in a pole-to-equator luminosity gradient that influences MASCARA-4 b's transit light curve and irradiation.}
\label{fig:teff}
\end{figure}

\begin{figure}[htbp]
\includegraphics[width=0.47\textwidth]{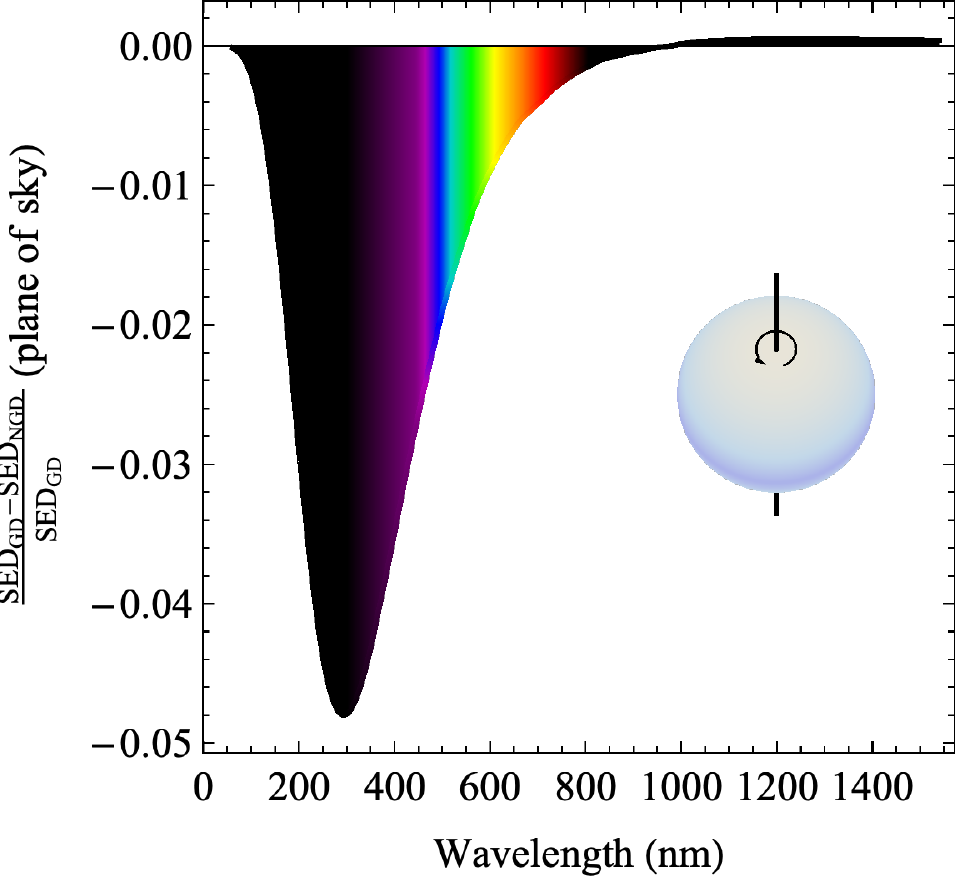} 
\caption{\footnotesize Rapid stellar rotation affects HD 85628's observable SED in two ways. First, its rotation induces an oblate stellar shape, increasing the size of the projected disk in the plane of the sky. Second, its gravity-darkened luminosity gradient causes the star to appear a bit cooler. Ultimately, these effects produce an SED that is shifted slightly down in the ultraviolet and visible and shifted very slightly up in the infrared compared to a slow rotator of equivalent size and temperature. This plot illustrates the difference between a gravity-darkened and traditional SED as seen in the plane of the sky using our measured and assumed stellar parameters. The inset figure shows HD 85628's sky-projected viewing geometry, with the color gradient representing the gravity-darkening gradient.}
\label{fig:SED}
\end{figure} 

Transit light curves have been modeled with gravity-darkening for a handful of planetary systems \citep[e.g.,][]{2011ApJS..197...10B,szabo2012spin,ahlers2014,ahlers2015spin,barnes2015probable,masuda2015spin,ahlers2019dealing,zhou2019two}. We follow the approach developed in \citet{barnes2009transit}, which used the Levenburg-Marquardt $\chi^2$ minimization routine to fit for bulk system parameters, stellar inclination, and projected stellar obliquity. We calculate parameter uncertainties from the covariance matrix stemming from the Levenburg-Marquardt routine, incorporating uncertainties in limb-darkening and gravity-darkening and using $\psi$ from \citet{dorval2019mascara} as an initial guess for our fit. Figure \ref{fig:angles} defines our orbit geometry parameters. 

The gravity-darkening technique constrains the true alignment of a planet, but cannot distinguish between a prograde or retrograde orbit. \citet{dorval2019mascara} previously determined MASCARA-4 b to be in a retrograde configuration; we therefore assume a retrograde orbit in our model, resulting in a single value for the planet's true alignment angle.

To model gravity-darkening, we use a previous constraint of HD 85628's $v\sin(i)$ from spectroscopy \citep{dorval2019mascara} and model the star's oblateness using the Darwin-Radau relation \citep[e.g.,][]{barnes2003measuring}. The star's high rotational velocity near its equator lessens its surface gravity, which changes its effective surface temperature as,

\begin{equation}
T_\mathrm{eff}=T_\mathrm{pole}\left(\frac{g_\mathrm{eff}}{g_\mathrm{pole}}\right)^\beta
\label{eq:gravdark}
\end{equation}

where $T_\mathrm{pole}$ and $g_\mathrm{pole}$ are the effective temperature and surface gravity at the star's poles, $T_\mathrm{eff}$ and $g_\mathrm{eff}$ are at any point on the star's surface, and $\beta$ is the gravity-darkening exponent that sets the strength of the temperature change across the stellar surface. The von Zeipel theorem that leads to Equation \ref{eq:gravdark} \citep{von1924radiative} predicts $\beta=0.25$, assuming a blackbody spectrum. However, subsequent theoretical and observational works have determined that $\beta$ is often below 0.25 due to thin convective envelopes that can suppress the gravity-darkening effect \citep[e.g.,][]{kervella2005gravitational,2007Sci...317..342M,lara2011gravity}. 

The gravity-darkening exponent $\beta$ is a difficult parameter to determine observationally for a given system; therefore we adapt $\beta=0.23^{+0.01}_{-0.02}$ from \citet{lara2011gravity}, which assumes that the star's energy flux is a divergence-free vector antiparallel to the effective gravity and matches the few available observations of gravity-darkening derived by interferometry \citep{2007Sci...317..342M,zhao2009imaging,che2011colder,jones2015ages}. We show HD 85628's effective temperature as a function of latitude in Figure \ref{fig:teff} and gravity-darkening's effect on the star's sky-projected spectral energy distribution (SED) in Figure \ref{fig:SED}.

Our transit model includes quadratic limb darkening with constants adapted from \citet{claret2017limb}. Using prior constraints of  the star's surface gravity ($\log(g)=4.0\pm0.5$), stellar effective temperature ($T_\mathrm{eff}=7800\pm200$ K), and solar metallicity ([Fe/H]$\sim$0), we use VizieR's limb-darkening tool\footnote{\url{http://vizier.u-strasbg.fr/viz-bin/VizieR-3?-source=J/A\%2bA/600/A30/tableab}} \citep{claret2017limb} and adapt $a=0.240^{+0.018}_{-0.005}$ and $b=0.245^{+0.03}_{-0.012}$ as HD 85628's quadratic limb-darkening coefficients for \tess's bandpass. We note that holding $\beta$ and limb-darkening parameters within an assumed range decreases the calculated uncertainty of the stellar inclination angle. See Table \ref{table:star} for a full list of stellar parameters. We show our best-fit results in \S\ref{sec:results.tess}. 

\renewcommand{\arraystretch}{1.2}
\begin{table*}[htbp]
\centering
\begin{tabular}{l l c r}
\hline \hline {\bf Parameter} & {\bf Description} & {\bf Value} & {\bf Source} \\ \hline
$P$ & orbital period (days) & $2.82406\pm0.00003$ &  \citet{dorval2019mascara} \\
$T_\mathrm{eff}$ & stellar effective temperature (K)  & $7800\pm200$ & \citet{dorval2019mascara} \\
$M_\star$ & stellar mass ($M_\odot$) & $1.75\pm0.05$ & \citet{dorval2019mascara} \\
$R_\star$ & stellar radius ($R_\odot$) & $1.92\pm0.11$ & \citet{dorval2019mascara} \\
$\log(g)$ & stellar surface gravity & $4.10\pm 0.05$ & \citet{dorval2019mascara} \\
{[Fe/H]} & metallicity & $\sim0$ & \citet{dorval2019mascara} \\
$v\sin(i)$ & sky-projected rotational velocity (km/s) & $46.5\pm1.0$ & \citet{dorval2019mascara} \\
$a$ & first limb-darkening term & $0.240^{+0.018}_{-0.005}$ & \citet{claret2017limb} \\
$b$ &  second limb-darkening term & $0.245^{+0.03}_{-0.012}$ & \citet{claret2017limb} \\ 
$\beta$ & gravity-darkening exponent & $0.23^{+0.01}_{-0.02}$ & \citet{lara2011gravity}\\ \hline
\end{tabular}
\caption{\footnotesize Previously-reported or assumed system parameters.}
\label{table:star}
\end{table*}

\subsection{Secondary Eclipse}
We report a detection of MASCARA-4 b's secondary eclipse in \tess's phase-folded FFI photometry. We fit the secondary eclipse and estimate $e\cos(\omega)$ using the primary and secondary transit times and $e\sin(\omega)$ using the primary and secondary transit durations, following \citet{charbonneau2005detection}. This approach gives only a weak constraint on $e\sin(\omega)$, so our analysis does not yield meaningful values for $e$ and $\omega$ individually. We constrain $e\cos(\omega)=-0.014\pm0.01$, indicating that MASCARA-4 b's orbit is slightly elliptical. Table \ref{table:bestfit} lists other relevant secondary eclipse parameters.

\subsection{Gravity-Darkened Insolation} \label{sec:methods.insolation}
\citet{ahlers2016gravity} first showed that planets in misaligned configurations around rapid rotators can receive unique insolations due to the star's asymmetric luminosity. In such a scenario, the planet varies in exposure to the host star's hot poles and cool equator, which can affect the planet's equatorial temperature and its incident XUV flux. Additionally, the star's projected disk as seen by the planet changes in size and peak emission throughout the orbit. 

We model MASCARA-4 b's gravity-darkened insolation following \citet{ahlers2016gravity}. We use quadratic limb-darkening and gravity-darkening parameters from our best-fit photometric model (Tables \ref{table:star} and \ref{table:bestfit}). We show the results of our model in \S\ref{sec:results.insolation}.

\section{Results} \label{sec:results}

\subsection{\emph{TESS} Photometry} \label{sec:results.tess}
We phase-fold and fit MASCARA-4 b's 15 transits observed by \tess~using the gravity-darkening transit model \citep{barnes2009transit} to determine the planet's spin-orbit angle. Our best-fit results in general agree with \citet{dorval2019mascara}. Most notably, our gravity-darkened fit reproduces the tight constraints on MASCARA-4 b's impact parameter and orbital alignment that \citet{dorval2019mascara} measured via Doppler tomography. We give our full list of measured parameters in Table \ref{table:bestfit}.

\begin{figure}
\includegraphics[width=0.47\textwidth]{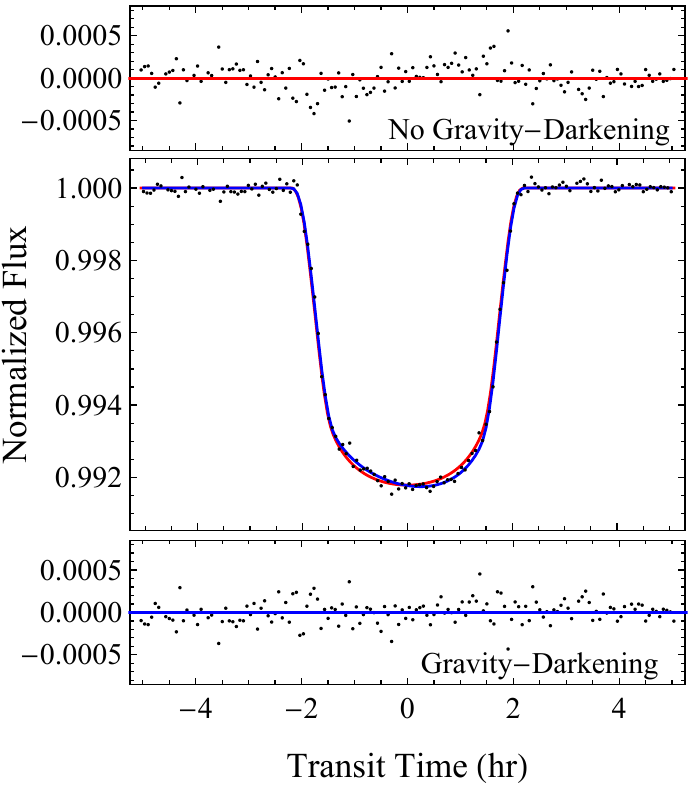}
\caption{\footnotesize MASCARA-4 b's \tess~light curve shows a clear left/right asymmetry due to its host star's gravity-darkened surface. The planet begins its transit near the star's dim equator and moves toward the star's bright pole, thus yielding a greater transit depth during egress. The above figure shows our best-fit model with and without gravity-darkening (blue and red, respectively). The gravity-darkening signal is evident in the top residual, in which a traditional best-fit model cannot resolve the star's asymmetry.}
\label{fig:fit}
\end{figure}

We test our best-fit gravity-darkened model against a traditional transit fit. Gravity-darkening better characterizes the slightly asymmetric shape of the transit light curve, confirming that the host star HD 85628 indeed has an asymmetrically luminous surface. Figure \ref{fig:fit} shows both models.

Our model takes advantage of the star's asymmetry to measure both the stellar inclination $i_\star$ and the projected stellar obliquity $\psi$. Together with the planet's orbital inclination $i$ (i.e., its impact parameter), we obtain a constraint of the planet's true spin-orbit angle $\varphi$ via,

\begin{equation}
\cos(\varphi) = \sin(\psi)\cos(i) + \cos(\psi)\sin(i)\cos(\lambda)
\label{eq:alignment}
\end{equation} 

We measure MASCARA-4 b to be in a nearly-polar orbit with $104^{\circ+7^\circ}_{-13^\circ}$ (see Figure \ref{fig:angles} for a representation of MASCARA-4 b's orbit geometry). We discuss possible mechanisms for causing misalignment in this system in \S\ref{sec:discussion.migration}.

\begin{figure}
    \centering
    \includegraphics[width=0.49\textwidth]{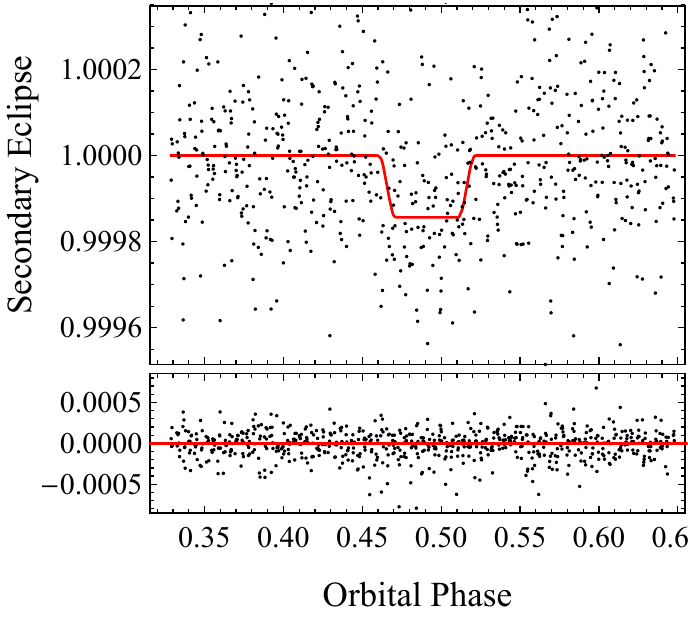}
    \caption{MASCARA-4 b's secondary eclipse occurs at $0.491\pm0.007$ orbital phase, indicating a slightly eccentric orbit. We measure an eclipse depth of $130\pm20$ ppm. We list measured eclipse parameters in Table \ref{table:bestfit} and discuss the eclipse depth in \S\ref{sec:results.tess}.}
    \label{fig:sec}
\end{figure}

\renewcommand{\arraystretch}{1.2}
\begin{table*}[htbp]
\centering
\begin{tabular}{l l c c c}
 \hline \hline {\bf Parameter} & {\bf Description} & {\bf G-Dark} & {\bf No G-Dark} & {\bf Dorval et al. (2019)} \\ \hline
$\chi^2_\mathrm{red}$ & goodness of fit & 1.092 & 1.4830  & 1.43 \\
$R_\star$ & polar stellar radius $(R_\odot)$ & $1.79\pm0.04$ & --- & $1.92\pm0.11$  \\
$R_\mathrm{p}$ & planet radius $(R_\mathrm{Jup})$ & $1.48\pm0.05$ & --- & $1.53^{+0.07}_{-0.04}$  \\
$R_\mathrm{p}/R_\star$ & radii ratio & $0.083\pm0.005$ & $0.086\pm0.003$ & $0.080^{+0.006}_{-0.005}$ \\
$T_0$ & transit epoch (BJD-2457000)  & $1573.5971\pm0.0003$ & $1573.5975\pm0.0002$ & $1505.817\pm0.003$\\
$b$ & impact parameter & $0.33\pm0.05$  & $0.36\pm0.05$  & $0.34\pm0.03$ \\
$i$ & orbital inclination (deg) & $86.7\pm0.5$  & $86.4\pm0.6$ & $88.50\pm0.01$ \\
$i_\star$ & stellar inclination (deg) &  $-63^{+10}_{-7}$ & --- & --- \\
$\psi$ & projected stellar obliquity (deg) & $244\pm 15$ & --- & $244.9^{+2.7}_{-3.6}$\\
$\varphi$ &  spin-orbit angle (deg) & $104^{+7}_{-13}$  & --- & --- \\
$\Omega_\star$ & stellar rotation period (hr) & $21^{+8}_{-7}$ & --- & ---  \\\
$\zeta$ & stellar oblateness  & $0.028\pm0.009$ & --- & --- \\
$T_\mathrm{sec}$ &secondary epoch (BJD-2457000) & --- & $1572.185\pm0.008$ & --- \\ 
$\delta F_\mathrm{sec}$ & secondary eclipse depth (ppm) & --- & $130\pm20$ & --- \\
$e\cos(\omega)$ & eccentricity & --- & $-0.014\pm0.01$ & --- \\
$e\sin(\omega)$ & eccentricity & --- & $0.032\pm0.065$ & --- \\
\hline
\end{tabular}
\caption{\footnotesize Best-fit parameters of MASCARA-4 b with and without gravity-darkening. Our gravity-darkened model more accurately resolves the ingress/egress asymmetry in transit depth seen in Figure \ref{fig:fit}, which reflects our better $\chi^2_\mathrm{red}$. We list previously-found values for MASCARA-4 b from \citet{dorval2019mascara}.}
\label{table:bestfit}
\end{table*}

We detect a $130\pm20$ ppm secondary eclipse in MASCARA-4 b's phase-folded \tess~photometry at $0.491\pm0.007$ orbital phase (Figure \ref{fig:sec}). The orbital phase is very near the midpoint between transits, indicating a slightly eccentric orbit. We calculate $e\cos(\omega)$ and $e\sin(\omega)$ for MASCARA-4 b following \citet{charbonneau2005detection}.

The secondary eclipse depth requires some explanation. The host star HD 85628's effective surface temperature is $7800\pm200$ K, so only $\sim23$\% of the star's emission falls within \tess's bandpass of $\sim600-1100$ nm. If the planet's emission were purely thermal with no reflected light, we could estimate MASCARA-4 b's equilibrium temperature at $3700\pm100$ K based on the eclipse depth. However, our best-fit results rule out this temperature as unphysical -- MASCARA-4 b simply cannot be that hot. Based on the star's effective temperature and radius and the planet's orbital period, we calculate the planet's equilibrium temperature to be near 1900 K for zero albedo; therefore, the secondary eclipse depth cannot be explained by emitted light alone. 

The secondary eclipse is likely being influenced by reflected light, which indicates a significantly non-zero bond albedo value. Our results are consistent with recent findings that some hot Jupiters posesess high-albedo clouds \citep[e.g.,][]{demory2011high,demory2013inference,parmentier2016transitions,barstow2016consistent}. High-precision follow-up observations of MASCARA-4 b's secondary eclipse could better determine the planet's equilibrium temperature and albedo, and could determine the existence of clouds. Given the high uncertainty in the secondary eclipse depth, we estimate MASCARA-4 b's equilibrium temperature at $\sim1900$ K based on the best-fit results of the primary transit and our gravity-darkening insolation model rather than the secondary eclipse.

\subsection{Insolation}\label{sec:results.insolation}
We simulate MASCARA-4 b's insolation accounting for rapid stellar rotation using our orbital configuration results from \S\ref{sec:results.tess}. Two effects result from rapid rotation that can affect insolation: the star's distorted shape, and its gravity-darkened surface. HD 85628 is slightly oblate and varies in effective temperature by $\sim220$ K between its pole and equator, which affects its local luminosity by $\sim12$\%.

\begin{figure*}
\begin{tabular}{r l}
\includegraphics[width=0.47\textwidth]{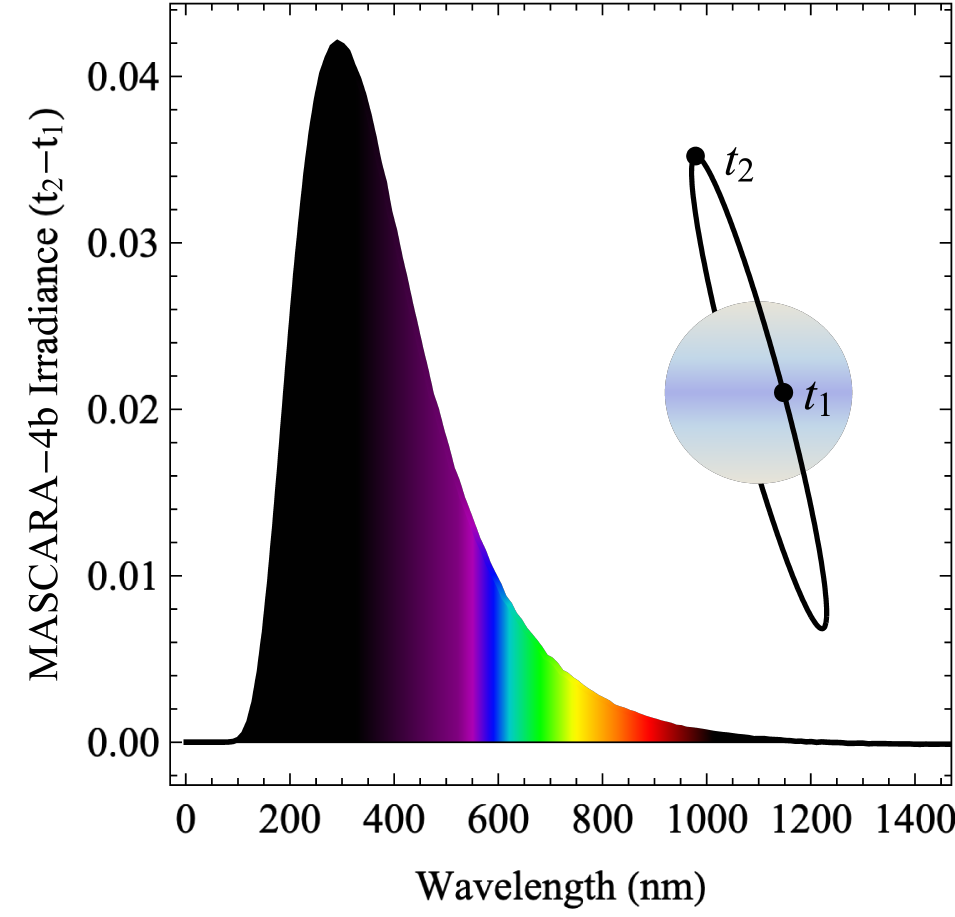} & \includegraphics[width=0.47\textwidth]{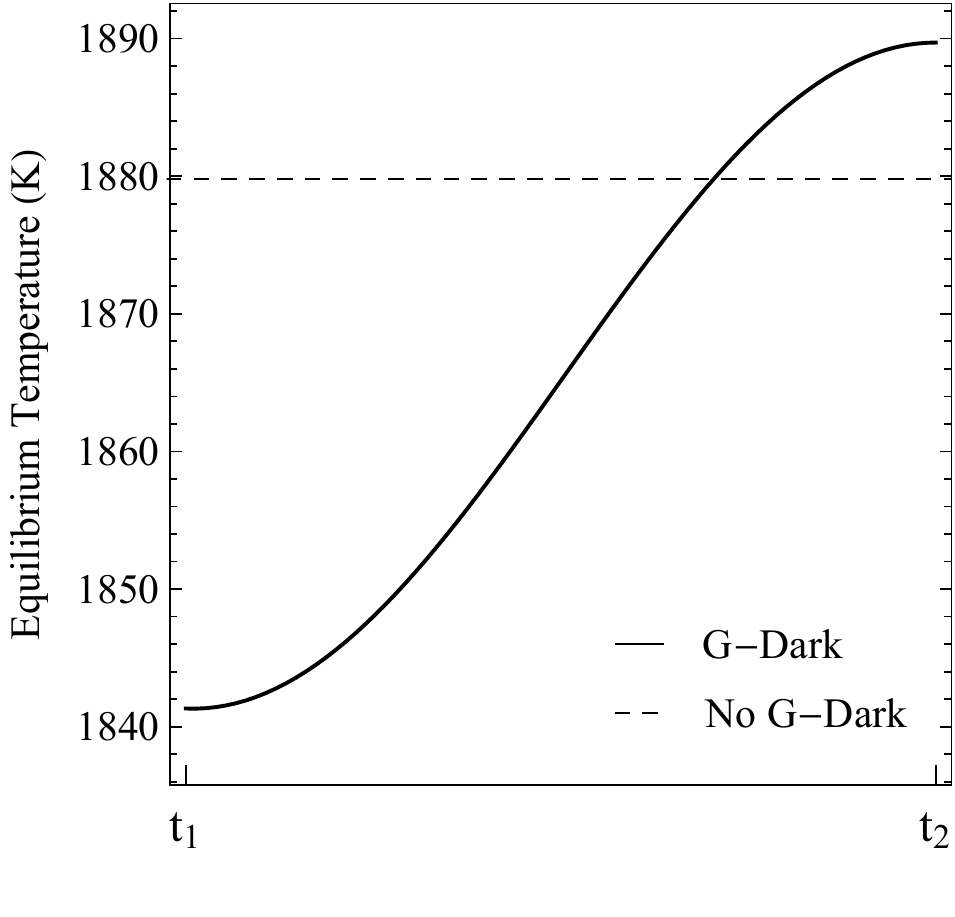} \\
\end{tabular}
\caption{\footnotesize MASCARA-4 b's irradiance changes throughout its orbit as it varies in exposure to its host star's hot poles and dim equator. The left figure shows the normalized difference between MASCARA-4 b's irradiance when residing in the star's equatorial plane ($t_1$) and when most exposed to the north stellar pole ($t_2$). At $t_2$, incident XUV flux is $\sim4$\% more intense. The right figure shows MASCARA-4 b's changing equilibrium temperature (assuming a bond albedo of 0) from $t_1$ to $t_2$ as well as the planet's theoretical equilibrium temperature when not accounting for rapid stellar rotation. We calculate a lower equilibrium temperature than derived in \citet{dorval2019mascara} because of our smaller best-fit stellar radius.}
\label{fig:irradiance}
\end{figure*}

When MASCARA-4 b resides near HD 85628's equatorial plane, the planet sees a slightly smaller projected disk and a slightly cooler, redder stellar surface. As it moves out of the star's equatorial plane, its exposure to one of the star's hot poles increases and the projected disk increases in size, increasing MASCARA-4 b's overall received flux. We show MASCARA-4 b's irradiance as a function of its orbital phase and its effects on the planet's theoretical equilibrium temperature in Figure \ref{fig:irradiance}. 

HD 85628's peak emission is near the border between visible and ultraviolet light; therefore, the largest relative change in MASCARA-4 b's irradiance is in the near ultraviolet. Overall, the effect of gravity-darkening on MASCARA-4 b's insolation is relatively weak because HD 85628's rotation period of $21^{+8}_{-7}$ hours is not all that fast compared to other A-type stars such as the well-known rapid rotators Vega (12.5 hr) \citep{peterson2006vega} or Altair (9 hours) \citep{2007Sci...317..342M}. However, ultraviolet light changing by several percent throughout the planet's orbit could have significant effects on its photochemistry and atmospheric processes.

We calculate that MASCARA-4 b's equilibrium temperature varies between 1840 K and 1890 K assuming a bond albedo of 0 based on our gravity-darkening model. These values represent simple equilibrium temperature calculations based on the star's apparent luminosity at each point of the orbit, following \citet{ahlers2016gravity}. Realistically MASCARA-4 b's thermal inertia would prevent such dramatic temperature changes on the planet as a whole, but its varying insolation can force the upper atmosphere significantly out of thermal equilibrium. Following Equation 4 from \citet{komacek2017atmospheric}, we estimate MASCARA-4 b's radiative timescale to be $\sim1$ day at $100$ mbar and $\sim10$ days at 1 bar. Therefore, the upper atmosphere of MASCARA-4 b is likely changing in temperature dramatically throughout the planet's 2.82 day orbit due to its host star's gravity-darkened surface. This effect may produce strong zonal winds that could vary in intensity with the varying received stellar flux, and that could match or exceed the fast wind speeds observed on other hot Jupiters \citep[e.g.,][]{snellen2010orbital,louden2015spatially,brogi2016rotation}

As shown in \S\ref{sec:results.tess}, MASCARA-4 b's secondary eclipse indicates that the planet is reflecting a significant amount of light, suggesting that the equilibrium temperature in Figure \ref{fig:irradiance} is inaccurate. We perform this analysis not to constrain MASCARA-4 b's true equilibrium temperature, but rather to demonstrate how gravity-darkening can influence a planet's insolation. We conclude that MASCARA-4 b's upper atmosphere likely changes in temperature significantly throughout its 2.82 day orbit and that its secondary eclipse depth implies a non-zero bond albedo. More robust constraints on MASCARA-4 b's equilibrium temperature and atmospheric processes are outside the scope of this project.

\section{Discussion}\label{sec:discussion}
As one of the hottest planets discovered to date, and as a planet residing in a 2.82-day polar orbit, MASCARA-4 b's dynamic formation history and current-day environment make it an excellent laboratory both for understanding hot Jupiters and for understanding planet formation around high-mass stars. In the following subsections we discuss possible migration scenarios for MASCARA-4 b, the effect of gravity-darkening on its current-day insolation, and future work to be done on the system. We also discuss the synergies between the gravity-darkening technique and Doppler tomography.

\subsection{Possible Migration Scenarios}\label{sec:discussion.migration}
The traditional nebular hypothesis predicts that MASCARA-4 b should reside beyond HD 85628's water ice line near the system's invariable plane; however, the planet is currently in a nearly-polar 2.82-day orbit. It therefore likely migrated inward during or after its formation, and some dynamic mechanism likely caused its orbit to tilt out of alignment. \citet{dorval2019mascara} identified a K/M stellar companion with projected separation of 740 AU, which could have played a significant role in the formation and migration of this system.

Several hypotheses have been postulated for how spin-orbit misalignment occurs. In general, they encompass three basic scenarios. One idea is that an outside body torques the system's protoplanetary disk out of alignment, and the planet forms inside the misaligned plane. \citet{batygin2012primordial} and others \citep{batygin2013magnetic,lai2014star,jensen2014misaligned} demonstrated that a stellar companion can torque a disk out of the formation plane, resulting in planets already misaligned when they form. \citet{batygin2012primordial} and \citet{zanazzi2018effects} demonstrated that precession of protoplanetary disks can lead to stellar obliquity angles greater than $90^\circ$. Similarly, \citet{bate2010chaotic} and \citet{fielding2015turbulent} showed that a wide range of stellar obliquities can occur when the star forms in a turbulent environment, which may have played a role in MASCARA-4 b's misalignment. 

Another possibility is that the host star's rotation axis torques out of alignment. In such a scenario, any planets ostensibly remain in their formation plane, and the star instead misaligns from the system. \citet{2012ApJ...758L...6R} and \citet{rogers2013internal} show that angular momentum transport in massive stars can torque  a star's envelope, resulting in a large apparent stellar obliquity. Such a process may be detectable via asteroseismic analysis; however, following \citet{ahlers2018lasr} we do not find any evidence of stellar pulsations in MASCARA-4 b's \tess~photometry. 

The third general idea for explaining spin-orbit misalignment is that some mechanism misaligned the planet after formation, which encompasses a wide variety of concepts. Kozai-Lidov resonance involves bodies exchanging angular momentum by driving up inclinations and eccentricities, which could explain MASCARA-4 b's polar orbit \citep{fabrycky2007shrinking}.  \citet{storch2014chaotic} demonstrated that Lidov-Kozai resonance can also cause a star's rotation axis to evolve chaotically, similarly producing spin-orbit misalignment. Additionally, spin-orbit misalignment can occur through secular interactions \citep{naoz2011hot} or violent scattering events \citep{morton2011discerning}. Unknown additional bodies in the system or HD 85628's stellar companion may have driven MASCARA-4 b through one or more of these scenarios.

Several theories exist to explain the inward migration of hot Jupiters, but given MASCARA-4 b's polar orbit, we posit high-eccentricity migration \citep[e.g.,][]{petrovich2015hot,mustill2015destruction} as a likely cause of inward migration in this system. A dynamic event such as Lidov-Kozai resonance or scattering could have raised both MASCARA-4 b's eccentricity and inclination (possibly torquing HD 85628's obliquity as well), and then the planet could have recircularized via tidal dissipation, maintaining its high inclination. Ultimately, determining the cause of misalignment is beyond the scope of this work; future projects studying the dynamic behavior of this system could better-constrain its migration history.

\subsection{Gravity-Darkened Seasons}\label{sec:discussion.seasons}
We show in \S\ref{sec:results.insolation} that MASCARA-4 b receives a varying irradiance due the star's gravity-darkened surface. Throughout its orbit, MASCARA-4 b's received flux varies by 4\% in the ultraviolet and slightly less in the visible. While such a variation would have an enormous impact on an Earth-like climate, it likely does not produce a detectable change in MASCARA-4 b's overall heat transport, winds, or cloud distribution. Similarly, the theoretical change in equilibrium temperature of $\sim50$ K likely causes dynamic atmospheric processes unlike anything seen in our solar system, but would likely not be measurable via phase curve.

The effect of gravity-darkening on a planet's insolation can be compared to the insolation of a planet with an eccentric orbit. In both scenarios the planets receive varying amounts of flux throughout their year, which can drastically impact climate. However, the frequency of changing flux is twice per orbit for gravity-darkening versus once per orbit in eccentricity. Additionally, the effects of gravity-darkening are chromatic (with the largest flux changes typically occurring in the near ultraviolet), whereas eccentricity is achromatic. Gravity-darkening likely plays a more significant role than eccentricity for the insolation of planets such as MASCARA-4 b because hot Jupiter orbits are typically nearly circular.

It is worth noting that the gravity-darkening effect on MASCARA-4 b is quite weak compared with many systems. For example, the hottest-known planet to date, KELT-9 b \citep{gaudi2017giant}, orbits an oblate A0 star that likely varies by more than a thousand Kelvin between its poles and equator. KELT-9 b's orbital configuration is very similar to MASCARA-4 b's, but the gravity-darkening effect on KELT-9 b's insolation is much stronger because its host star rotates much more rapidly. Similarly, Kepler-462 b \citep{ahlers2015spin} orbits a rapidly-rotating star, and with an orbital period of 85 days, its response to gravity darkening is likely quite large because it goes through much longer exposures to the star's hot poles and cool equator. While gravity-darkening may not be all that impactful for MASCARA-4 b's seasons, it likely has substantial effects on a large number of planets orbiting high-mass stars.

\subsection{Gravity-Darkening vs Doppler Tomography}
Gravity-darkening and Doppler tomography complement each other in a number of ways. Both techniques constrain an exoplanet's spin-orbit geometry, but do so in ways that work synergistically with one another. \citet{dorval2019mascara} previously analyzed MASCARA-4 b with Doppler tomography and we apply gravity-darkening in this work, giving MASCARA-4 b one of the most robustly determined orbit geometries to date.

Gravity-darkening is advantageous over Doppler tomography in three ways. First, gravity-darkening constrains the true spin-orbit angle --- an angle otherwise very difficult to obtain --- by measuring both the host star's projected obliquity and inclination. Second, it relies almost entirely on high-precision transit photometry, which space telescopes like \emph{Kepler} and \tess~provide in abundance. Third, it provides constraints on the host star's rotation period, gravity-darkened surface, and oblateness. However, gravity-darkening is computationally expensive and difficult to model, and has only been applied to a handful of planets. Previous works have struggled to resolve the interdependence between the gravity-darkening exponent (Equation \ref{eq:gravdark}) and limb-darkening. To date, this work marks only the third occurrence where gravity-darkening results are confirmed by Doppler tomography \citep{zhou2019two,johnson2014misaligned,masuda2015spin}; further confirmation of gravity-darkening would strengthen the model's validity.

On the other hand, Doppler tomography is advantageous over gravity-darkening in three ways. First, the approach is well-understood and produces robust projected obliquity measurements. Second, it typically provides a tight constraint on the planet's impact parameter, which can be difficult to obtain via transit photometry. Third, Doppler tomography easily distinguishes between a prograde and retrograde transit, which gravity-darkening cannot do. In a prograde orbit, the planet first blocks light from the half of the star that rotates towards the observer, causing a net redshift. As the planet transits, it then covers the half of the star which orbits away from the observer, causing a net blueshift. This can be seen through cross-correlation functions of spectra taken during the transit as a dark shadow moving from $-v\sin(i)$ to $+v\sin(i)$ of the star \citep{cegla2016rossiter}. The exact opposite happens if the planet is retrograde. \citet{cegla2016rossiter} provides an overview of Doppler tomography with its advantages and weaknesses.

The best method for characterizing a planet with both Doppler tomography and gravity-darkening is therefore the approach adopted in \citet{dorval2019mascara} and this work: obtain constraints of projected obliquity (including prograde/retrograde transit orientation), impact parameter, and $v\sin(i)$ via Doppler tomography, and then apply the gained knowledge as priors for gravity-darkening. The combined approach yields a robust measurement of true spin-orbit angle with two independent measurements of the projected alignment, constraints on the host star's rotation period and asymmetry, and the bulk system parameters yielded from standard transit analysis. 

\subsection{Future Work}\label{sec:discussion.future}
MASCARA-4 b is a hot Jupiter in a 2.82-day polar orbit around a bright ($V_\mathrm{mag}=8.19$) AV3 star, making it an excellent target for further study via follow-up observations. While MASCARA-4 b is not in an environment quite as extreme as KELT-9 b, recent studies of KELT-9 b demonstrate just how exotic these ultra-hot Jupiters can be. For example, ground-based studies by \citet{Cauley2019} and \citet{Hoeijmakers2019} have revealed the presence of metals like magnesium, iron, titanium in the extended atmosphere of KELT-9 b. With a bright host star and a mass and radius similar to KELT-9 b, MASCARA-4 b is a promising target for similar atmospheric detections using both high-resolution ground-based spectrographs and space-based facilities like the \emph{Hubble Space Telescope} and the \emph{James Webb Space Telescope}. Atmospheric characterization of these misaligned ultra-hot Jupiters provides constraints on the composition of their atmospheres that may in turn reveal clues to their formation history.

In its primary mission, \tess~is expected to observe approximately 397,000 stars of sufficient mass to be rapid rotators\footnote{from the \emph{TESS} Web Viewing Tool} and should find $\sim2000$ planets around A/F stars -- many of which will have spin-orbit misaligned orbits \citep{barclay2018revised}. Using MASCARA-4 b as a test case, we can estimate that a large fraction of these newly-discovered planets will make excellent targets for the gravity-darkening technique. First, an estimated 92 of those planets' host stars will have a brighter \tess~magnitude than HD 85628 ($m_\mathrm{\emph{TESS}}=8.047$). Second, approximately 530 of those \tess~discoveries will be observed in more than one sector, yielding impressive photometric precision. Third, the gravity-darkening signal on HD 85628 is relatively weak due to its somewhat unimpressive rotation rate. Many newly-discovered planets will transit host stars with significantly stronger gravity-darkening, making the signal easier to detect. With gravity-darkening easily detectable in MASCARA-4 b's transit light curve, it is reasonable to expect a prolific survey of gravity-darkened targets from \tess.   

\acknowledgements
This paper includes data collected by the TESS mission, which are publicly available from the Mikulski Archive for Space Telescopes (MAST) and produced by the Science Processing Operations Center (SPOC) at NASA Ames Research Center \citep{jenkinsSPOC2016}. Funding for the TESS mission is provided by NASA's Science Mission directorate. Resources supporting this work were provided by the NASA High-End Computing (HEC) Program through the NASA Advanced Supercomputing (NAS) Division at Ames Research Center for the production of the SPOC data products. J.P.A.’s research was supported by an appointment to the NASA Postdoctoral Program at the NASA Goddard Space Flight center, administered by Universities Space Research Association under contract with NASA. I.S. acknowledges funding from the European Research Council (ERC) under the European Union’s Horizon 2020 research and innovation program under grant agreement No 694513. 

\facilities{TESS}

\bibliography{MASCARA4b_paper}
\bibliographystyle{aasjournal}

\end{document}